\documentclass[letterpaper, 10pt, conference]{ieeeconf}  

\IEEEoverridecommandlockouts                              
\usepackage{amsmath,graphicx,epstopdf,array,color}
\usepackage{epsfig}
\usepackage{float}
\usepackage{graphics} 
\usepackage{mathptmx} 
\usepackage{times} 
\usepackage{amsmath} 
\usepackage{amssymb}  
\usepackage{bm}
\usepackage{fancyhdr}
\let\labelindent\relax
\usepackage{mathtools}
\usepackage[ruled]{algorithm2e}
\usepackage{enumerate}
\usepackage{easyReview}

\newtheorem{definition}{Definition}

\DeclareMathOperator*{\argmax}{argmax}

\allowdisplaybreaks 

\title{\LARGE \bf Control invariant set enhanced reinforcement learning for process control: improved sampling efficiency and guaranteed stability}

\author{Song Bo, Xunyuan Yin, Jinfeng Liu
\thanks{
Song Bo and Jinfeng Liu are with the Department of Chemical \& Materials Engineering, University of Alberta, Edmonton Canada, T6G 1H9. Xunyuan Yin is with the School of Chemistry, Chemical Engineering and Biotechnology, Nanyang Technological University, 637459, Singapore. Corresponding author: J. Liu. Email: jinfeng@ualberta.ca.
}}

\begin{document}

\maketitle
\thispagestyle{empty}
\pagestyle{empty}

\begin{abstract}
Reinforcement learning (RL) is an area of significant research interest, and safe RL in particular is attracting attention due to its ability to handle safety-driven constraints that are crucial for real-world applications of RL algorithms. This work proposes a novel approach to RL training, called control invariant set (CIS) enhanced RL, which leverages the benefits of CIS to improve stability guarantees and sampling efficiency. The approach consists of two learning stages: offline and online. In the offline stage, CIS is incorporated into the reward design, initial state sampling, and state reset procedures. In the online stage, RL is retrained whenever the state is outside of CIS, which serves as a stability criterion. A backup table that utilizes the explicit form of CIS is obtained to ensure the online stability. To evaluate the proposed approach, we apply it to a simulated chemical reactor. The results show a significant improvement in sampling efficiency during offline training and closed-loop stability in the online implementation. 
\end{abstract}

\section{Introduction}

In process control, model predictive control (MPC) is a standard approach to optimal control. It is formulated as a constraint optimization problem in which the safety constraints are taken into account explicitly \cite{mayne2000constrained}. However, for large-scale systems, MPC may suffer from high computational complexity. Reinforcement learning (RL), as one main component of machine learning, provides an alternative to MPC for optimal control and can shift the complex optimization calculations to offline training based on a model \cite{sutton2018reinforcement}. It has gain significant attention in different industries, (for example, game \cite{mnih2015human}, finance \cite{singh2022reinforcement}, energy \cite{chen2022reinforcement}) for decision-making and control purposes. 

RL is a class of optimal control algorithms that enables machines to learn an optimal policy (closed-loop control law), by maximizing future rewards through repetitive interactions with the environment~\cite{sutton2018reinforcement}. It uses a trial-and-error approach to interact with the environment, allowing it to learn and find the optimal policy even in the absence of prior knowledge of the process. In addition, the consideration of future rewards in RL ensures that current decisions are beneficial in the long run. However, the standard RL approach does not incorporate safety constraints in its design and does not guarantee closed-loop stability, which limits its use in real-world applications \cite{garcia2015comprehensive}. To address these challenges, safe RL algorithms have been developed, which explicitly consider safety constraints during training and ensure closed-loop stability in the learned policy.

Safe RL is a class of RL algorithms that aims to achieve a stable closed-loop control system by taking safety constraints into account. Different approaches to designing safe RL have been summarized in the literature \cite{garcia2015comprehensive, osinenko2022reinforcement, gu2022review}. One existing approach is to consider the constrained Markov decision process (CMDP), in which a cost function is used to penalize undesired actions, transforming the original optimization problem into a new problem where both reward maximization and action cost minimization are required \cite{law2005risk, gehring2013smart}. This approach increases the probability of safe actions but without guarantees. Another approach is to use MPC to guide the RL algorithm by treating MPC as parameterized value or policy neural networks \cite{zanon2019practical, gros2021reinforcement}. However, these approaches still require solving the MPC optimization problem recursively, resulting in a high computational burden.

Sampling efficiency is {another critical issue that has limited real-world applications of RL} \cite{yu2018towards}. In the context of stability, an RL algorithm with low sampling efficiency requires a significant number of agent-environment interactions to achieve a stable and optimal control policy, leading to prohibitively high costs in its application. To address this challenge, an intuitive solution is to allow the agent to interact only with the controllable states of the environment. By doing so, the agent can find a policy that maintains the system within the controllable state space, achieving the stability guarantee while eliminating unnecessary interactions with uncontrollable states and improving sampling efficiency. Unfortunately, such environments are generally unavailable.

In the field of control theory, it is widely acknowledged that control invariant sets (CIS) play a crucial role in ensuring the stability of a control system \cite{decardi2022robust}. These sets characterize the states within which a feedback control law is always available to maintain the system within the set~\cite{blanchini1999set}. Incorporating the concept of CIS in RL is expected to improve the stability guarantee and sampling efficiency by restricting the agent's interactions with the system to only controllable states. This way, the agent can find a policy that maintains the system within the CIS and achieves stability with fewer interactions with the environment. The concept of CIS has indeed been adopted in RL design to achieve closed-loop stability. The main idea is to filter or project risky actions to safe ones, typically by adding a stand-alone safety filter after the learned RL policy \cite{alshiekh2018safe, gros2020safe, li2020robust, tabas2022computationally}. However, because the filter only considers safety, the optimality that the controller is trying to achieve is not always preserved. To address this, \cite{gros2020safe} proposes embedding CIS in the last layer of the RL policy network to enable back-to-back training and achieve both safety and optimality. Due to the challenge of obtaining a CIS for a general nonlinear system, researchers have shifted their focus towards implicit methods that utilize control barrier functions (CBF), Hamilton-Jacobi (HJ) reachability analysis, and safe backup controllers to define safety constraints and design filters indirectly \cite{brunke2022safe}. 

Though the above algorithms take into account safety in the training of the RL, the sampling efficiency remains as a critical issue. Moreover, it is worth noting that these studies combining CIS and RL have been conducted mainly in robotics, and limited research has been carried out in process control. In the realm of process control, process systems are in general highly nonlinear, tightly interconnected and of large-scale. These features present challenges in applying the above mentioned CBF and HJ analysis based algorithms. Furthermore, control problems beyond set-point or reference trajectory tracking, such as zone tracking \cite{decardi2022robust,zhang2020} and economic optimization \cite{ellis2014}, are common in process control. These control objectives add additional complexities that make the above mentioned approaches difficult to use. 

While the construction of a CIS is not a trivial task, various methods have been developed in the past decade. For example, algorithms for constructing or approximating the CIS for constrained linear systems \cite{rungger2017computing, rakovic2005invariant} and general nonlinear systems \cite{homer2017constrained, fiacchini2010computation} have been proposed. Graph-based approaches to find the outer and inner approximations of robust CIS of nonlinear systems has also been developed \cite{decardi2021computing}. Over the past couple years, data-driven approaches have also been used to find the invariant sets of nonlinear systems, which approximate invariant sets using neural nets \cite{chen2021learning, bonzanini2022scalable}. These approaches facilitate the study of safe RL that utilizes a CIS explicitly.

The above considerations motivate us to study the explicit integration of RL and CIS for process control, where the CIS can serve as a state space for the RL agent to explore, safely. Minimal modification to the RL algorithms is required, while the reward function design can incorporate both economic or zone tracking objectives which are common in process control. Specifically, in this work, the CIS of a nonlinear process is assumed to be available. Then, a two-stage CIS enhanced RL is proposed to improve the sampling efficiency and guarantee the stability. The first stage involves offline training with a process model and the CIS. Due to the potential disastrous consequences of failed process control, the use of a model to pre-train the RL offline can provide a significant amount of data with strong temporal correlation and broad coverage of various scenarios. The introduction of CIS has the potential to narrow down the state space, reduce the training dataset size, and provide guidance on agent exploration. However, exhaustive training cannot guarantee that the RL agent has encountered every scenario, which may result in instability in online implementation. Hence, the second online implementation stage involves online learning when the safety constraint is violated. A new control implementation strategy is proposed to ensure closed-loop stability. The proposed approach is applied to a chemical reactor to demonstrate its applicability and efficiency.

\section{Preliminaries}

\subsection{System description}

In this work, we consider a class of nonlinear processes that can be described by the following discrete-time state-space form:
\begin{equation} \label{eqn:nonlinear}
    x_{k+1}=f(x_k, u_k)
\end{equation}
where $x \in X \subseteq \mathbb{R}^{n}$ and $u \in U \subseteq \mathbb{R}^{m}$ denote the state and the input {vectors} of the process with $X$ and $U$ being the physical constraints on $x$ and $u$, $f: \mathbb{R}^{n} \times \mathbb{R}^{m} \rightarrow \mathbb{R}^{n}$ is a nonlinear function mapping the present state to the state at the next time instant, $k$ represents the time index.

\subsection{Reinforcement learning}

Reinforcement learning broadly represents {the class of }data-driven learning algorithms in which an agent is trying to learn a closed-loop policy $\pi(u|x)$, a conditional probability of prescribing $u$ at given state $x$, by interacting with the environment. 
The Markov {decision process} (MDP) is utilized to formulate the environment. The environment receives the action prescribed by the agent and provides the resulting reward and state of the system back to the RL agent. {The} state transition dynamics of the MDP is shown below:
\begin{equation} \label{eqn:state_transition}
    {P}(r_{k+1}, x_{k+1}|x_k, u_k)
\end{equation}
where $x_k$ denotes the current state of the environment, $u_k=\pi(x_k)$ is the action prescribed by the agent based on the learned policy, $r_{k+1}$ represents the reward used for criticizing the action, $x_{k+1}$ represents the state sampled at the next sampling time instant, {${P}$} denotes the conditional probability of the state transition. 

As in \cite{sutton2018reinforcement}, the RL problem can be formulated as the following:
\begin{equation}
		\pi^* = \argmax_{\pi}{\mathbb{E}_{\pi}[G_k|x_k, u_k]}
\end{equation}
where $G_k$ denotes the return accumulating the reward $r$ in long run. The optimal policy $\pi^*$ is found when the expected return following the such policy is maximized.

In this work, the environment dynamics that describe the transition from $x_k$ to $x_{k+1}$ is represented by the nonlinear system of Eq.~(\ref{eqn:nonlinear}). Note that in system~(\ref{eqn:nonlinear}), uncertainty is not considered for brevity.

\subsection{Control invariant sets}

A control invariant set of a system is a set of states in which the system can stay inside all the time by following a feedback control law. The definition of the control invariant set is given below:
\begin{definition}[c.f. \cite{blanchini1999set}]\label{def:cis}
The set $R \subseteq X$ is said to be a control invariant set for system~(\ref{eqn:nonlinear}) if for any $x_k\in R$, there exists an input $u_k\in U$ such that $x_{k+1}\in R$.
\end{definition}

In the control literature, CISs play an important role in ensuring the stability of the closed-loop systems. For example, CISs are commonly {used} in MPC designs as a terminal constraint for achieving guaranteed stability and feasibility \cite{mayne2000constrained, cannon2003nonlinear, decardi2022robust}. 

\subsection{Problem formulation}

Standard RL does not consider the safety constraints which obstructs its application{. Also, conventional RL typically requires} a significant amount of data for training. A CIS inherently provides the region of operation that is stable and may be integrated in RL offline and online training to ensure closed-loop stability. In addition, the introduction of the CIS is possible to narrow down the state space, to reduce the training dataset size, and to provide guidance on agent exploration.

The objective of this work is to propose a CIS enhanced RL and its training method to guarantee the closed-loop stability and improve the sampling efficiency during RL training, by incorporating the CIS knowledge into RL. The RL optimization can be described as follows:
\begin{equation}
\begin{aligned}
    \pi^* = \argmax_{\pi} \quad & {\mathbb{E}_{\pi}[G_k|x_k, u_k]}\\
    \textrm{s.t.} \quad & x_{k}\in R \\
    & u_k\in U
\end{aligned}
\end{equation}
where the state constraint as well as the input constraint are considered. 

\section{Proposed approach} \label{sec:proposed_procedure}

In this section, we present the proposed CIS enhanced RL. In the proposed approach, it includes both offline training and online training. The first step is to train the RL with the CIS information offline to achieve a near-optimal policy. The incorporation of the CIS in the offline training can significantly improve its sampling efficiency {since the amount of data needed for training is reduced; this} will be demonstrated in the simulations. While the CIS is used in offline training, the offline trained policy does not guarantee the closed-loop stability. In order to ensure the closed-loop stability, the RL should further be trained during its online implementation. A new control implementation strategy is proposed to ensure that the applied control actions ensure the closed-loop stability. Figure~\ref{fig:proposed_alg} illustrates the proposed approach {and the difference between the proposed approach and the fundamental RL are highlighted in blue. The details of the two steps are explained below. 

\begin{figure}
\centering
\includegraphics[width=\columnwidth]{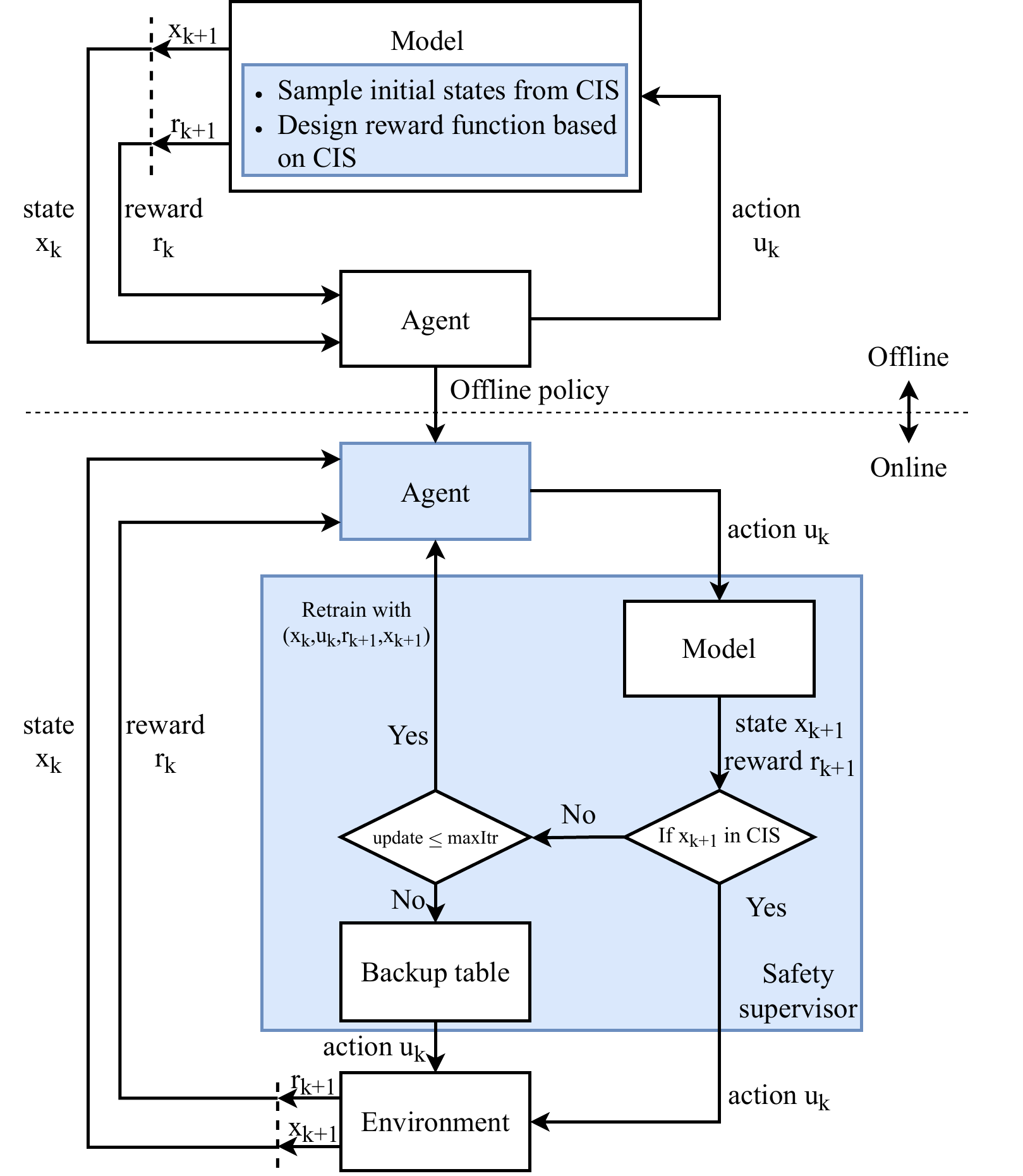}
\caption{Flow diagram of the proposed RL training approach.}
\label{fig:proposed_alg} 
\end{figure}

\subsection{Offline training} \label{sec:proposed_train}

It is assumed that a CIS of system~\eqref{eqn:nonlinear}, $R$, is available before the training of the RL. It is preferred that the CIS is the maximum CIS of \eqref{eqn:nonlinear} within the physical constraint $X$, which ensures that the RL can explore within the biggest feasible and stable operating region. Meanwhile, we note that the maximum CIS is not a requirement in the proposed approach, any CIS can be used in the proposed approach. 

In the proposed RL offline training, the system model \eqref{eqn:nonlinear} is used in the training of the RL. In this work, we focus on demonstrating the concept of the proposed approach and assume that there is no model uncertainty. The CIS information is used in two ways. First, the CIS is used to penalize the RL agent when it drives the system state outside of the CIS. This can be achieved by designing the reward function appropriately. A discrete reward function shown in Eq.~\eqref{eqn:reward_general} may be used. 
\begin{equation} \label{eqn:reward_general}
    r(x_{k},u_k)= 
    \begin{cases}
    r_1,& \text{if } x_{k+1} \in R\\
    r_2              & \text{otherwise}
    \end{cases}
\end{equation}
where $R\subset X$ denotes the CIS of system~\eqref{eqn:nonlinear} used in the RL training, $r_1 \in \mathbb{R}$ and $r_2 \in \mathbb{R}$ denote the reward values associated with the prescribed action $u_k$ based on the current state $x_k$. Note that $r_1$ should be greater than $r_2$ ($r_1 > r_2$) to guide the RL to prescribe control actions that can maintain the system state within the CIS. Specifically, at the time instant $k$, the system is at state $x_k$. When the RL prescribes the control action $u_k$, it is sent to the model and the model will propagate into the next time instant and obtain $x_{k+1}$. If $x_{k+1}$ is within the CIS ($x_{k+1}\in R$), the RL will receive a higher reward $r_1$; if $x_{k+1}$ is outside of the CIS ($x_{k+1}\notin R$), indicating the prescribed action resulting an unstable operation, the RL receives a relatively lower reward (or penalty) $r_2$. 

Moreover, the CIS is used for initial state sampling in RL offline training. In RL offline training, an RL needs to sample the initial state of the system randomly for many times. Typically, the RL is restricted to sample the initial condition within the physical constraint set $X$. In the proposed approach, instead of using $X$, we propose to sample the initial state $x_0$ of the system within the CIS $R$. If the CIS is the maximum CIS, and if the system starts from an initial state outside of the CIS, the RL is not able to stabilize the system and drive the system back into the CIS. Such a case indeed does not provide too much information for learning a policy that ensures the stability. If the system starts within the CIS, then RL is able to find a control action to stabilize the system and learn the optimal and non-optimal actions based on the reward function~{(\ref{eqn:reward_general})}. Therefore, the sampling efficiency can be improved by restricting the RL to sample initial states within the CIS. This will be demonstrated in the simulations in the next section.

Another technique we propose in the offline training is to reset the state to its previous value when the state is outside of the CIS. Assume that at time $k$, $x_k\in R$. The RL agent prescribes a control action $u_k$ and drives the system state to be outside of the CIS; that is, $x_{k+1}\notin R$. In such a case, the RL will get a lower reward $r_2$ according to \eqref{eqn:reward_general}. Since once the system state is outside of the CIS, there is no control action that can drive the system back to the CIS (if the CIS is the maximum one within $X$) and the system becomes unstable, the interaction between the RL and the system will not further bring much useful information towards learning the optimal control law. Therefore, we propose to reset the state to its previous value; that is, set $x_{k+1}=x_k$ and then continue the training process. By implementing this state resetting technique, the RL learns from this failure experience and will get second or more chances to learn at the same state $x_k$ towards the stable and optimal policy.

\subsection{Online training and stability guaranteed implementation strategy} \label{sec:proposed_stability}

After offline training, the RL learns a policy and the RL policy is saved as a pre-calculated controller for online implementation. However, due to the sampling nature of the RL training, it is impossible for the offline trained RL agent to encounter all situations. Therefore, the offline learned policy may not guarantee the closed-loop stability. To address this issue, we propose to implement the RL policy using a stability guaranteed strategy and to further train the RL policy online when a new situation is encountered. As shown in Figure~\ref{fig:proposed_alg}, a safety supervisor is placed in between the RL agent and the environment. The detailed description of the safety supervisor is shown below.

Let us consider the current sampling time is $k$. With the state feedback $x_k$, the RL prescribes the control action $u_k$ according to the learned policy. Based on $u_k$ and the system model~\eqref{eqn:nonlinear}, the state $x_{k+1}$ at the next sampling time is predicted. If the predicted state $x_{k+1}$ is within the CIS ($x_k\in R$), then the control action $u_k$ is actually applied to the system; if the predicted state $x_{k+1}$ is outside of the CIS ($x_k\notin R$), then the RL is switched again to the training mode and the policy is updated with the new interaction experience $(x_k,u_k,r_{k+1},x_{k+1})$. The updated policy will prescribe the updated action based on the state $x_k$ again. Unless the new policy guarantees that the predicted state $x_{k+1}$ is within the CIS, the agent will keep learning at the current state $x_{k}$ until a pre-determined maximum number of iterations (\text{$maxItr$}) is reached. This online training/updating approach can significantly enhance the safety of the RL. However, the closed-loop stability is still not guranteed. It is possible that within the maximum iteration $maxItr$, the online updating of the RL may not converge to a stable action (the RL cannot find a stable action for the particular state $x_k$ within $maxItr$ online updates). It should be pointed out that when $maxItr$ is very large, the online training is expected to find a safe action for every state it encounters since the CIS provides the guarantee of the existence of a safe action for all the states within it. 

To address the above issue and to guarantee the closed-loop stability, we propose the use of a backup table to save the safe actions for the states such that the above online training fail to find a safe action within $maxItr$ iterations. The safe actions can either be found using another stabilizing but not optimal control law, or by sampling the control action space randomly, or by leveraging the information contained in the explicit form of the CIS. Note that in some CIS construction algorithms \cite{decardi2022graph}, the corresponding safe action space for each state is also found, which can be taken advantage of in creating the backup table. This approach provides a safety guarantee for the RL.

The proposed stability guaranteed implementation strategy and online training is summarized in the following algorithm:

\begin{algorithm}
	\caption{Safety supervisor in online implementation for stability guarantee}\label{alg:two}
	\KwIn{$x_k$, $k$, $maxItr$}
	\KwOut{Safe $u_k$}
	$notSafe \gets True$, $update=1$\;
	\While{notSafe}{
		Calculate $u_k$ at $x_k$ based on the learned RL policy\;
		Based on the model and $u_k$, predict $x_{k+1}$\;
		\eIf{$x_{k+1} \in R$}{$notSafe \gets False$\;}
			{	
			\eIf{$update \leq maxItr$}{Update RL policy with $(x_k,u_k,r_{k+1},x_{k+1})$\;
			$update \leftarrow update+1$\;}
			{Get safe action $u_k$ from the backup table\; $notSafe \gets False$\;}		
			}
	}
	Apply $u_k$ to system~\eqref{eqn:nonlinear} and obtain $x_{k+1}$\;
	Reinitialize the algorithm with $k\leftarrow k+1$
\end{algorithm}

In the algorithm, the parameter $maxItr$ can be defined by the user to balance between the computational complexity and optimality of the RL agent. A larger value will allow the agent to be trained on a state for a longer time, which can potentially result in a better safety performance. However, this comes at the cost of increased online computational complexity. On the other hand, a small value, or even zero, will ensure stability and online implementation feasibility given that the backup table is designed well. However, it may not achieve optimal performance when relying on the backup plan for safe actions, because the selection among safe actions at the current state does not consider the optimality. One more factor to consider in picking, $maxItr$, is the sampling time of the system (time interval between two discrete time instants). It should be ensured that within one sampling time a stabilizing control action can be prescribed which indeed limits the maximum value of $maxItr$. 

Note that in the above discussion, the primary focus was on maintaining the system state within the CIS ($x_k\in R$), which is also the concept of stability considered in this work. Since $R\subset X$, it also ensures that the state constraint ($x_k\in X$) is satisfied. A rigorous stability proof of the proposed design is omitted for brevity. One interesting feature of the proposed approach is that maintaining $x_k\in R$ is handled through the incorporation of the CIS in the offline training and the online implementation. This provides flexibility in the proposed approach to incorporate other control objectives such as set-point tracking, zone tracking or economic optimization in the design of the reward function $r$. 

Note also that in this work, model uncertainty is not considered. When model uncertainty presents, the proposed design can be adapted to account for uncertainty, for example, by considering a robust CIS.

\section{Simulation results and discussion}

\subsection{Process description}

In order to study the sampling efficiency and the closed-loop stability guarantee of the proposed RL training and implementation approach, the application of the proposed approach to the control of a continuously stirred tank reactor (CSTR) is considered in this section. The reaction happening inside of the reactor is an irreversible and exothermic reaction with first-order reaction rate. The CSTR is also installed with a cooling jacket outside of it for maintaining the temperature of the reaction mixture. The mathematical model {contains} two nonlinear ordinary differential equations with the following representation {\cite{decardi2022robust}}:
\begin{eqnarray*} 
    && \frac{dc_A}{dt}=\frac{q}{V}(c_{Af}-c_A)-k_{0}exp(-\frac{E}{RT})c_{A} \\
     &&   \frac{dT}{dt} = \frac{q}{V}(T_f-T)+\frac{-\Delta H}{\rho c_p}k_{0}exp(-\frac{E}{RT})c_{A}+\frac{UA}{V\rho c_{p}}(T_c-T)
\end{eqnarray*}
where $c_A$ $(mol/L)$ and $T$ $(K)$ denote the concentration of the reactant and the temperature inside of the reaction mixture, respectively. $c_{Af}$ $(mol/L)$ and $T_f$ $(L)$ represent the concentration of the reactant and temperature of the inlet stream. $T_c$ $(K)$ is the temperature of the coolant stream used for cooling the reactor temperature. The remaining parameters are summarized in {Table~\ref{tbl:parameters}}. The parameter $q$ is the inlet and outlet flow {rate} of the reactor, $V$ is the volume of the reaction mixture, $k_0$ is the pre-exponential factor of the Arrhenius rate constant, $E$ represents the activation energy required by the reaction, $R$ denotes the universal gas constant, $\Delta H$ is the change of the enthalpy used for approximating the change of internal energy of the reaction, $\rho$ is the density of the reaction mixture, $c_p$ denotes the specific heat capacity of the reaction mixture, $UA$ is the heat transfer coefficient between the reactor and the cooling jacket.

\begin{table}
\centering 
\caption{Parameters of the CSTR model} 
\begin{tabular}{ccc} 
 \hline 
 Parameter &       Unit &      Value \\
 \hline
         $q$ &      $L/min$ &        $100$ \\

         $V$ &          $L$ &        $100$ \\

         $k_0$ &      $min^{-1}$ &        $7.2 \times 10^{10}$    \\

         $E/R$ &      $K$      &        $8750.0$    \\

         $- \Delta H$ &      $J/mol$ &   $5.0 \times 10^{4}$ \\

         $\rho$ &        $g/L$ &       $1000.0$ \\

         $c_p$ &       $J/gK$ &      $0.239$ \\

         $UA$ &     $J/minK$       &    $5.0 \times 10^4$        \\
         
		 $c_{Af}$   & $mol/L$ & $1$\\
		 $T_{f}$    & $K$ & $350$\\
		\hline
\end{tabular}  \label{tbl:parameters}
\end{table}

In the following closed-loop control problem study, $c_A$ and $T$ are the two states of the system; $T_c$ is the manipulated variable. They are subject to the following physical constraints: 
\begin{eqnarray}
    0.0\leq c_A \leq 1.0 \label{cstrcon1}\\
    345.0\leq T \leq355.0 \label{cstrcon2}\\
    285.0 \leq T_c \leq 315.0
\end{eqnarray}

The control objective is to train an RL policy to maintain a stable operation of the CSTR such that the two states are maintained within the physical constraints shown in \eqref{cstrcon1} and \eqref{cstrcon2} all the time.

The maximum CIS of the CSTR is calculated using the graph-based algorithm developed in \cite{decardi2021computing}. The physical constraints and the calculated maximum CIS over the state space are shown in Figure~\ref{fig:representation_of_cis}. According to Figure~\ref{fig:representation_of_cis}, the CIS spans over the entire temperature space and shrinks over the concentration space. Hence, if the concentration of the reactant is lower than the minimum value of $c_A$ in CIS (the top left point of CIS), no matter how $T_c$ is manipulated by the controller, the system becomes unstable. The same observation applies to when the system has a concentration that is higher than the maximum value of $c_A$ in CIS (bottom {right} point of CIS). Since the calculated CIS is the maximum one, when the system is outside of the CIS, there is no feedback control law that is able to bring the system back into CIS again and the system state will eventually diverge. This implies that the physical constraints will be violated. 

\begin{figure}
\centering
\includegraphics[width=\columnwidth]{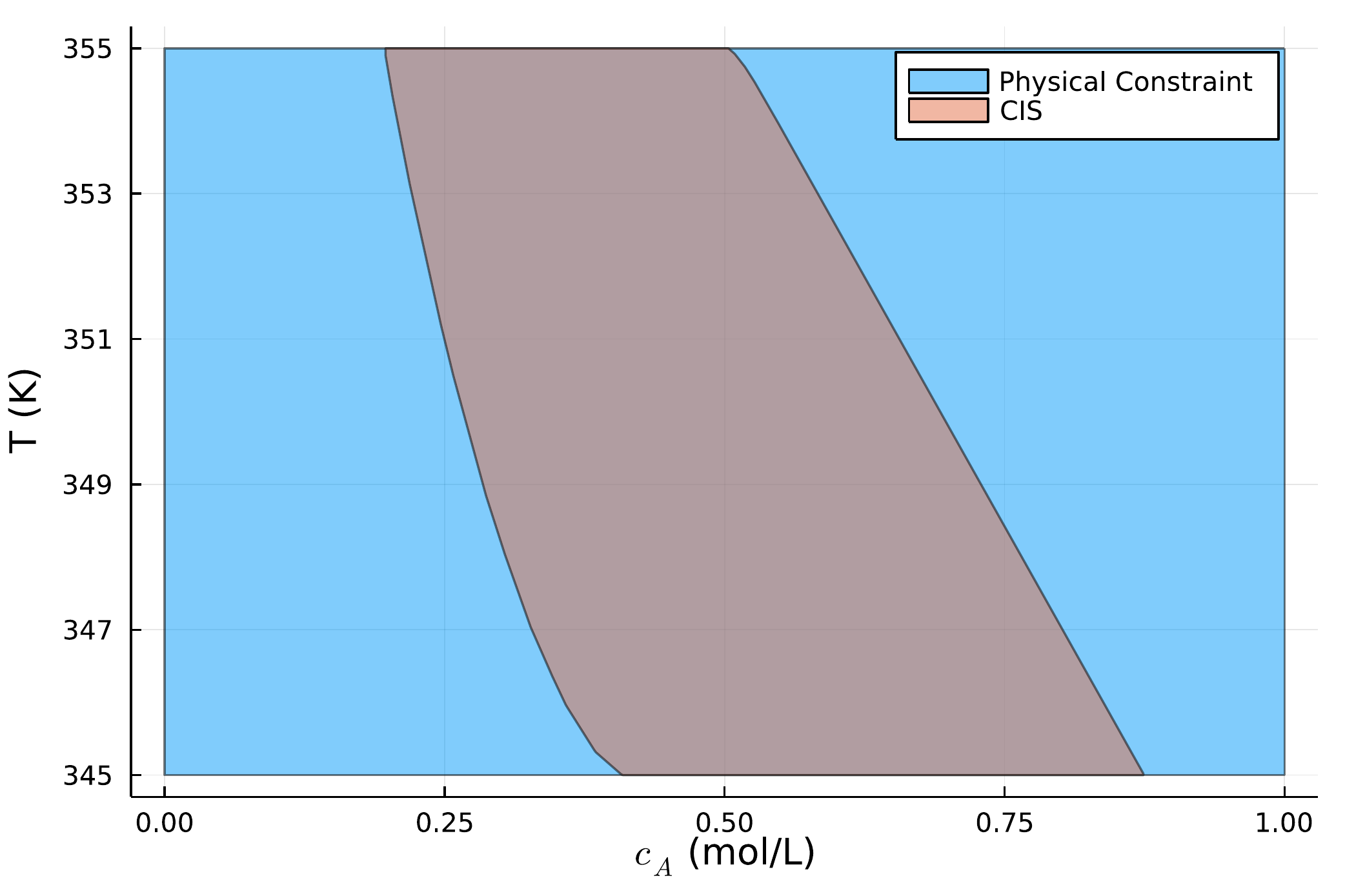}
\caption{The physical constraints and the maximum CIS of the CSTR.}
\label{fig:representation_of_cis} 
\end{figure}

\subsection{RL training setup and results} \label{sec:rl_train_setup}

In the training of the proposed RL, the maximum CIS is used. The proximal policy optimization (PPO) is used as the optimization algorithm in RL training. Even though, the nature of the CSTR system is a continuous process, the RL agent is trained under the episodic setting. The reason is that when the system reaches its steady state as time goes by, the collected data, the steady state transition, will not provide any new information to the agent. Though the exploration-exploitation embedded in PPO algorithm may prescribe actions enhancing the exploration, meaning at the given steady state PPO will not prescribe the corresponding steady input, the probability of the exploration becomes extremely low. Hence, in order to favor the exploration, once the agent interacting with the environment for a user-defined number of time steps, the episode is terminated and a new initial state is sampled. During the experiment, 10,000 episodes and 200 steps per episode {were} used to train the RL agent. The batch size was defined as 10 episodes, meaning that the RL agent would be updated only {when all} 10 episodes {were} finished and the RL would learn from the data of the 10 episodes at once. The learning rate {was} defined as $10^{-4}$, discounted factor {was} $0.99$. It was noticed that the offline training took overall 2,000,000 steps which might be computational expensive. However, the trained policy can be implemented online as a pre-calculated function which will require less online computation resources.}

The consideration of CIS in the proposed RL training setup is reflected in two steps, the sampling of the initial states and the design of reward function. According to Section~\ref{sec:proposed_train}, since the largest CIS is known, the initial state has to be inside of CIS to ensure the following states get the chance to be inside of CIS. Hence, all 10,000 initial states for 10,000 episodes were sampled within the CIS. In addition, because it {was} undesired for the system to enter the space outside of CIS, a discrete reward function was proposed as the following:
{\begin{equation} \label{eqn:reward}
    r(x_k,u_k)= 
    \begin{cases}
    10,000,& \text{if } x_{k+1} \in CIS\\
    -1,000              & \text{otherwise}
    \end{cases}
\end{equation}}

\begin{figure}
\centering
\includegraphics[width=0.9\columnwidth]{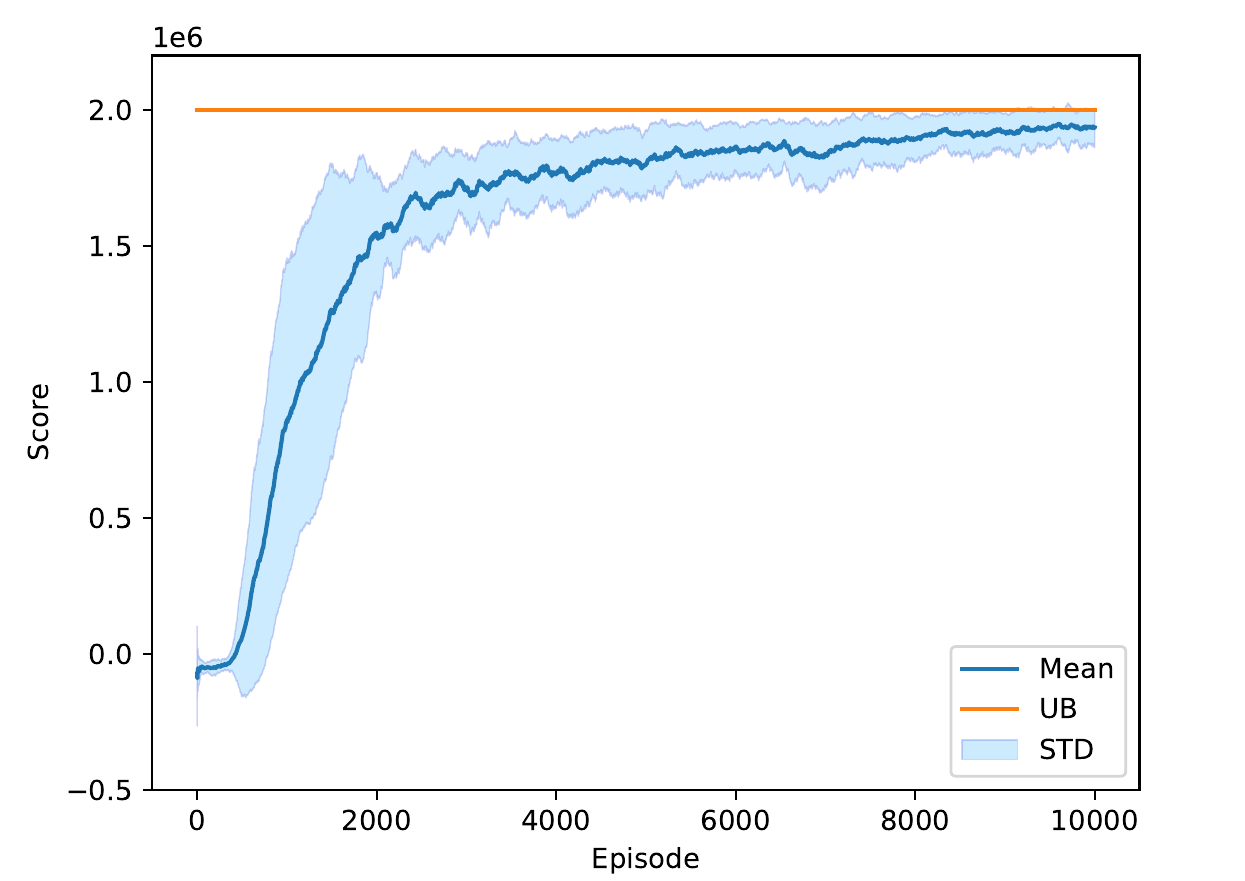}
\caption{Average training score of proposed RL design}
\label{fig:train_reward} 
\end{figure}

Based on aforementioned RL training setup, 20 offline training were executed in parallel and the learning performance was calculated as the average of performance over 20 training. The average training reward plot, representing the learning performance, is shown in Figure~\ref{fig:train_reward}. The orange horizontal line represents the maximum score each episode can achieve if the RL agent maintain the system within the CIS for all 200 steps{; the maximum score for each episode is $200 \times 10,000 = 2\times 10^6$}. The mean curve was calculated based on all 20 training and blue shaded area shows the one standard deviation. {Please note, in order to smooth out the fluctuations of the scores among episodes, the running average, which recursively calculates the average of scores of past 100 episodes, was used to plot the figure.} As RL agent interacting with the environment for more episodes, the score of episode increased, meaning RL agent was able to learn from the training. From the beginning to around 2,000$^{th}$ episodes, the RL agents had a higher learning rate with a larger variance between 20 training. After that, the learning slowed down and gradually reached the plateau with a decreased variance.

\subsection{RL testing setup and results} \label{sec:rl_test_setup}

Since in the RL training approach proposed in Section~\ref{sec:proposed_procedure}, an RL policy is trained offline before interacting with the environment online, it enables us to test the performance of the offline RL policy. The policy was tested with the model for 10,000 episodes and each episodes lasted for 200 steps. During the tests, the reward value was collected for evaluation purpose and not for learning purpose. Note that in all these tests, the control actions prescribed by the RL agents {were} applied directly to the CSTR and the proposed stability guaranteed online implementation strategy {was} not used. 

Table~\ref{tbl:test_fail_rate} summarizes the failure rates of the 20 trained RL policies. The failure is defined as when the policy cannot maintain the process within the CIS in an episode. According to the table, Run 18 was able to achieve a 0.02\% of failure rate. In other words, out of 10,000 episodes, there were only 2 episodes that the agent was failed to keep them inside of the CIS. Run 12 has the highest failure rate of 13.90\%. Overall, the proposed CIS enhanced offline RL training was able to achieve an average of 8.42\% failure rate. 

\begin{table}
	\centering
	\caption{Failure rates of RLs in testing mode}
	\begin{tabular}{cccc} 
		\hline
		Run \# &      Failure rate (\%) & 	Run \# &      Failure rate (\%)\\
		\hline
		1 &      7.91 & 11&8.14\\  
		
		2 &     5.79 & 12&13.90\\
		
		3 &    3.41    & 13&13.88\\
		
		4 &     12.48     & 14&10.78\\
		
		5 &   2.25   & 15&12.92\\
		
		6 &  12.01     & 16&12.85\\  
		
		7 &    3.76   & 17&12.94\\
		
		8 &    4.21    & 18&0.02\\
		
		9 &     13.37     & 19&7.44\\
		
		10 &   4.39   & 20&6.04\\
		\hline
	\end{tabular}  \label{tbl:test_fail_rate}
\end{table}




\subsection{Study of stability guarantee} \label{sec:stability_guarantee}

After the offline RL training, one RL agent was saved and treated as the pre-calculated feedback controller. Then by following the algorithm proposed in Section~\ref{sec:proposed_stability}, the agent was implemented online. The agent interacted with the environment for 10,000 episodes and each episodes lasted for 200 steps. In order to examine the benefits of the proposed online implementation, RL Run 1 obtained from Section~\ref{sec:rl_train_setup} was picked, because a near average test performance {was observed}. 
	
First of all, with the proposed online implementation strategy, the agent was able to maintain the system within the CIS for all 10,000 episodes. This is expected since the proposed online implementation is stability guaranteed. 
	
Second, since the agent was retrained during its implementation, it was expected that the retrained RL agent would be able to achieve a better performance in terms of stability. Therefore, two agents, one from offline training and one from online implementation, were compared and tested on one set of 10,000 initial states. Since the same set of initial states were used, the RL agent obtained after offline training was able to achieve a 7.91\% of failure rate which is the same as the value shown in Table~\ref{tbl:test_fail_rate}. The RL agent obtained after online training was able to reach a 0.02\% failure rate. By comparing these values, it shows that the proposed online implementation not only is able to ensures the stability, but also obtain a better RL agent. 

\subsection{Study of sampling efficiency} \label{sec:sampling_efficiency}

In order to study and quantify the benefits brought by utilizing the CIS in RL offline training, the sampling efficiency was studied. The study was conducted by comparing the results shown in Section~\ref{sec:rl_train_setup} with an RL training setup that {did} not utilize the CIS information. Hence, in this RL without CIS training, 10,000 of initial states were sampled within the physical constraints and the reward condition was extended from CIS in Eq.~(\ref{eqn:reward}) to the whole physical constraint. In addition, the reset of the state was done when the system state was outside of the physical constraints. Other parameters remained the same. 

Since two RL training setups had different reward functions, it was impossible to compare their training plots directly. {Hence, they were tested by comparing the failure rates introduced in Section~\ref{sec:rl_test_setup}}. The same 10,000 initial states were tested. Table~\ref{tbl:sampling_efficiency} shows that the RL with CIS was able to achieve 8.42\% of failure rate over 10,000 episodes and RL without CIS could only achieve 34.30\%. Therefore, using the same amount of data in training, utilizing the knowledge of CIS facilitates the learning process. Then both RL agents were trained with 20,000, 30,000, 40,000 and 50,000 episodes, then tested with the same 10,000 initial states that were used before. The failure rate of RL with CIS had a minor improvement to 4.84\% and that of RL without CIS had a relative bigger improvement to 14.58\%. However, the failure rate of RL without CIS training using 50,000 episodes {(14.58\%)} was still higher than that of RL with CIS trained only using 10,000 episodes {(8.42\%)}. Hence, the utilization of CIS improves the sampling efficiency of RL training process.

\begin{table}
\centering
\caption{Failure rates of RLs trained with different sizes of dataset}
\begin{tabular}{ccc} 
 \hline
 Training dataset size &      RL with CIS &      RL without CIS \\
 \hline
         10,000 &      8.42\% &     34.30\%    \\  

         20,000 &          8.30\% &        25.84\% \\

         30,000 &      6.55\% &        19.83\%    \\

         40,000 &       5.52\%     &        16.56\%    \\

         50,000 &    4.84\%   &   14.58\% \\
 \hline
\end{tabular}  \label{tbl:sampling_efficiency}
\end{table}

\section{Concluding remarks}

A CIS enhanced RL training and online implementation approach was proposed to obtain stability guaranteed RL implementation. The offline training stage incorporated the CIS in reward function design, the initial state sampling and the state reset technique. A stability guaranteed online implementation strategy was proposed for the implementation of the offline trained RL and the RL {was} also retrained if a new situation {was} encountered. The approach was applied on a two-dimensional nonlinear CSTR system. The results showed that the offline training stage was able to {provide an agent with a lower failure rate as compared to RL without CIS}. {Also,} the sampling efficiency was significantly improved {as} CIS was utilized in offline training. The online implementation stage ensured the stability and resulted in a better RL agent in terms of maintaining system inside of CIS.

\end{document}